\documentclass[manuscript]{aastex}
\usepackage{rotating,graphicx}

\shorttitle{Radial Dependence of SEP Peak Intensities}

\shortauthors{He, Zhou, \& Wan}

\begin{document}

\title{Propagation of Solar Energetic Particles in Three-dimensional Interplanetary Magnetic Fields: Radial Dependence of Peak Intensities \\}

\author{H.-Q. He\altaffilmark{1}, G. Zhou\altaffilmark{1,2}, and W. Wan\altaffilmark{1}}

\altaffiltext{1}{Key Laboratory of Earth and Planetary Physics,
Institute of Geology and Geophysics, Chinese Academy of Sciences,
Beijing 100029, China; hqhe@mail.iggcas.ac.cn}

\altaffiltext{2}{College of Earth Sciences, University of Chinese
Academy of Sciences, Beijing 100049, China}

\begin{abstract}
A functional form $I_{max}(R)=kR^{-\alpha}$, where $R$ is the radial
distance of spacecraft, was usually used to model the radial
dependence of peak intensities $I_{max}(R)$ of solar energetic
particles (SEPs). In this work, the five-dimensional Fokker-Planck
transport equation incorporating perpendicular diffusion is
numerically solved to investigate the radial dependence of SEP peak
intensities. We consider two different scenarios for the
distribution of spacecraft fleet: (1) along the radial direction
line; (2) along the Parker magnetic field line. We find that the
index $\alpha$ in the above expression varies in a wide range,
primarily depending on the properties (e.g., location, coverage) of
SEP sources and on the longitudinal/latitudinal separations between
the sources and the magnetic footpoints of the observers.
Particularly, the situation that whether the magnetic footpoint of
the observer is located inside or outside the SEP source is a
crucial factor determining the values of index $\alpha$. A two-phase
phenomenon is found in the radial dependence of peak intensities.
The ``position" of the breakpoint (transition point/critical point)
is determined by the magnetic connection status of the observers.
This finding suggests that a very careful examination of magnetic
connection between SEP source and each spacecraft should be taken in
the observational studies. We obtain a lower limit of
$R^{-1.7\pm0.1}$ for empirically modelling the radial dependence of
SEP peak intensities. Our findings in this work can be used to
explain the majority of the previous multispacecraft survey results,
and especially to reconcile the different/conflicting empirical
values of index $\alpha$ in the literature.
\end{abstract}


\clearpage

\section{Introduction}
Solar energetic particles (SEPs), because of their radiation effects
and potential damages to space missions, have become a focus of
space physics and space weather research. In addition, to achieve a
better understanding of SEP propagation in interplanetary space is
helpful for us to unveil the physical mechanisms of transport and
acceleration processes of energetic charged particles including
cosmic rays in the universe, which is a long-standing and
fundamental problem in the fields of heliophysics, astrophysics, and
plasma physics. To estimate the potential impact of SEPs on space
probes, the knowledge of the radial dependence of SEP intensities
and fluences is practically required and theoretically meaningful,
especially for the forthcoming Solar Probe Plus and Solar Orbiter.
Due to its importance, the variation of SEP intensities with radial
distance has been recently investigated by a number of authors in
observational community and numerical modeling community.

In previous studies, a functional form $I_{max}(R)=kR^{-\alpha}$,
where $R$ is the heliocentric radial distance of the spacecraft, was
usually used to model the radial dependence of peak intensities
$I_{max}(R)$ of SEPs. \citet{McGuire1983} estimated that the
averaged peak intensity of the prompt component of the SEP events
observed between 0.3 and 1 AU decreases a factor of 20 per AU with
increasing radial distance $R$. \citet{Hamilton1988} and
\citet{Hamilton1990} utilized the spherically symmetric transport
model based on \citet{Parker1965} to investigate the radial
dependence of peak intensities in SEP events. They particularly
deduced that the peak intensities of $10-20$ MeV protons decrease
with increasing radial distance in a functional form
$R^{-3.3\pm0.4}$. \citet{Shea1988} used measurements of $10-70$ MeV
protons in SEP events from 1 to 5 AU to assemble a descriptive model
of solar particles in the heliosphere. \citet{Kallenrode1997} fitted
the time-intensity profiles and time-anisotropy profiles of $4-13$
MeV protons in 44 SEP events and inferred that the index $-\alpha$
varied in the wide range $[-5.5, 4.5]$, with a median value of $-2$.
\citet{Smart2003} summarized the recommendations on the
extrapolation of SEP fluxes and fluences from 1 AU, and indicated
that the radial dependencies of SEP fluxes have a range of power
indices. \citet{Rosenqvist2003} averaged the particle intensities
measured by the Helios spacecraft from $0.3$ to $1.0$ AU and deduced
that the radial dependence of SEP intensities varied with particle
energy and ranged from $R^{-0.77}$ for $>4$ MeV protons to $R^{1.0}$
for $>51$ MeV protons. \citet{Lario2006} used the SEP measurements
of IMP 8 and two Helios spacecraft and deduced that the radial
distributions of SEP events showed ensemble-averaged variation
ranging from $R^{-2.7}$ to $R^{-1.9}$ for $4-13$ and $27-37$ MeV
proton peak intensities, respectively. Recently, numerical modeling
of longitudinal and radial dependence of SEP fluxes and fluences has
been intensely used in the community to understand the transport
effects of SEPs and to predict the radiation environment at various
heliocentric distances
\citep{McKenna-Lawlor2005,Aran2005,Lario2007,Kozarev2010,He2011,He2015a,He2017,Rouillard2011,Lario2011}.
\citet{Verkhoglyadova2012} numerically solved the focused-diffusion
transport equation taking into account the effects of a traveling
shock and perpendicular diffusion, and suggested that the functional
dependence of SEP radial distribution is softer than $R^{-3}$ and
specifically is about $R^{-2.9}$ to $R^{-1.8}$ for $0.3-5$ MeV
particles. \citet{Lario2013} used simultaneous measurements of SEP
events by MESSENGER and spacecraft near 1 AU (e.g., STEREO-A,
STEREO-B, ACE) to determine the radial dependence of
near-relativistic electron intensities. They deduced that the
$71-112$ keV electron peak intensities in the prompt component of
the SEP events decreased with increasing radial distance $R$ in a
functional form $R^{-\alpha}$ with both $\alpha<3$ and $\alpha>3$
for specific events.

In this work, we numerically solve the five-dimensional
Fokker-Planck transport equation to systematically investigate the
radial dependence of SEP peak intensities in the inner heliosphere.
The numerical model includes essentially all the SEP transport
mechanisms, such as streaming along magnetic field lines, convection
with solar wind, magnetic focusing, adiabatic deceleration, parallel
diffusion along the magnetic field, and perpendicular diffusion
across the magnetic field. We model the radial dependence of SEP
peak intensities with a functional form $I_{max}(R)=kR^{-\alpha}$,
where $k$ is a constant, $R$ is the heliocentric radial distance,
and $\alpha$ is a power-law index. The main purpose of this work is
to determine the index $\alpha$ in various SEP event scenarios,
e.g., different particle energies, different diffusion coefficients
(both parallel and perpendicular), and different source properties.
We take into account two different styles for the alignment of the
fleet of spacecraft in the heliosphere: (1) along the radial
direction line; (2) along the nominal Parker magnetic field line. We
find that the index $\alpha$ varies in a wide range, primarily
depending on the properties (e.g., location, coverage) of SEP
sources and on the longitudinal/latitudinal separations between the
sources and the magnetic field line footpoints of the observers.
Particularly, the situation that whether the magnetic footpoint of
the observer is located inside (even very near or on the boundary of
source) or outside the SEP source is a dominant factor determining
the values of index $\alpha$. We find a two-phase phenomenon in the
radial dependence of SEP peak intensities. The location of the
breakpoint (transition point/critical point) is determined by the
status of magnetic connection of the observers. A lower limit of
$R^{-1.7\pm0.1}$ to the radial dependence of SEP peak intensities is
deduced in our numerical simulations. In addition, we find that the
index $\alpha$ does not strongly depend on the energies of particles
and the ratios of perpendicular to parallel diffusion coefficients,
provided that the values of the diffusion coefficients are
relatively reasonable. Our findings in this work can be used to
explain the majority of the previous survey results based on
multispacecraft observations, and particularly to reconcile the
different/conflicting empirical values of index $\alpha$ in the
literature.

This paper is structured as follows. In Section 2, we briefly
describe the five-dimensional Fokker-Planck transport equation and
the numerical method for solving it. We also illustrate the specific
physical scenarios of SEP transport modelling. In Section 3, we
present the numerical modelling results and discuss the radial
dependence of SEP peak intensities. Finally, we summarize our main
results in Section 4.

\section{Numerical Modelling Based on Fokker-Planck Transport Equation}
The five-dimensional Fokker-Planck transport equation that governs
the gyrophase-averaged distribution function $f(\textbf{x},\mu,p,t)$
of SEPs can be written as
\citep[e.g.,][]{Schlickeiser2002,Zhang2009,He2011,He2015b,Droge2014}
\begin{eqnarray}
{}&&\frac{\partial f}{\partial t}+\mu v\frac{\partial f}{\partial
z}+{\bf V}^{sw}\cdot\nabla f+\frac{dp}{dt}\frac{\partial f}{\partial
p}+\frac{d\mu}{dt}\frac{\partial f}{\partial \mu}  \nonumber\\
{}&&-\frac{\partial}{\partial\mu}\left(D_{\mu\mu}\frac{\partial
f}{\partial \mu}\right)-\frac{\partial}{\partial
x}\bigg(\kappa_{xx}\frac{\partial f}{\partial x}\bigg)
-\frac{\partial}{\partial y}\left(\kappa_{yy}\frac{\partial
f}{\partial y}\right)=Q({\bf x},p,t),  \label{transport-equation}
\end{eqnarray}
where $\textbf{x}$ is spatial location of particles, $z$ is the
coordinate along the magnetic field line, $p$ is momentum of
particles, $\mu$ is pitch-angle cosine of particles, $t$ is time,
$v$ is speed of particles, $\textbf{V}^{sw}$ is solar wind speed,
$\kappa_{xx}$ and $\kappa_{yy}$ are perpendicular diffusion
coefficients, and $Q$ is source term. The expression $dp/dt$ in
Equation (\ref{transport-equation}) represents effect of adiabatic
deceleration and can be written as
\begin{equation}
\frac{dp}{dt}=-p\left[\frac{1-\mu^2}{2}\left(\frac{\partial
V^{sw}_x}{\partial x}+\frac{\partial V^{sw}_y}{\partial
y}\right)+\mu^2\frac{\partial V^{sw}_z}{\partial z}\right].
\label{adiabatic-cooling}
\end{equation}
In addition, the expression $d\mu/dt$ includes magnetic focusing
effect and divergence of solar wind flows and can be written as
\begin{eqnarray}
\frac{d\mu}{dt}&=&\frac{1-\mu^2}{2}\left[-\frac{v}{B}\frac{\partial
B}{\partial z}+\mu \left(\frac{\partial V^{sw}_x}{\partial
x}+\frac{\partial V^{sw}_y}{\partial y}-2\frac{\partial
V^{sw}_z}{\partial z}\right)\right]     \nonumber\\
{}&=&\frac{1-\mu^2}{2}\left[\frac{v}{L}+\mu\left(\frac{\partial
V^{sw}_x}{\partial x}+\frac{\partial V^{sw}_y}{\partial
y}-2\frac{\partial V^{sw}_z}{\partial z}\right)\right],
\label{magnetic-focusing}
\end{eqnarray}
where $B$ is mean interplanetary magnetic field, and the magnetic
focusing length $L$ is defined by
$L=(\textbf{z}\cdot\bigtriangledown \ln B)^{-1}$.

The parallel mean free path $\lambda_{\parallel}$ in the diffusion
approximation regime can be written as
\begin{equation}
\lambda_{\parallel}=\frac{3v}{8}\int_{-1}^{+1}\frac{(1-\mu^{2})^{2}}{D_{\mu\mu}}d\mu.
\label{parallel-path}
\end{equation}
The radial mean free path can be accordingly written as
\begin{equation}
\lambda_{r}=\lambda_{\parallel}\cos^{2}\psi, \label{radial-path}
\end{equation}
where $\psi$ denotes the angle between local magnetic field
direction and radial direction. In addition, $\cos^{2}\psi$ in
Equation (\ref{radial-path}) can be written as
\begin{equation}
\cos^{2}\psi=(V^{sw})^2/\left((V^{sw})^2+\Omega^2R^2\sin^2\theta\right),
\label{cos2psi}
\end{equation}
where $\Omega$ is angular rotation speed of the Sun, $R$ is
heliocentric radial distance, and $\theta$ is colatitude.

In this work, we use a pitch-angle diffusion coefficient with the
functional form as \citep[e.g.,][]{Beeck1986,Zhang2009,He2011}
\begin{equation}
D_{\mu\mu}^{r}=D_{\mu\mu}/\cos^{2}\psi=D_{0}vR_{d}^{-1/3}\left(|\mu|^{q-1}+h\right)(1-\mu^{2}),
\label{diffusion-coefficient}
\end{equation}
where $D_{0}$ is a constant indicating strength of magnetic
turbulence, $R_{d}$ is rigidity of particles, $h$ is a constant used
to model particles' scattering ability through $90^{\circ}$
pitch-angle ($\mu=0$), and $q$ is a constant (chosen to be $5/3$ in
this work) related to magnetic turbulence's power spectrum in
inertial range.

We use time-backward Markov stochastic process technique to solve
the five-dimensional Fokker-Planck Equation
(\ref{transport-equation}). By using this technique, the
Fokker-Planck Equation (\ref{transport-equation}) can be easily
recast into five time-backward stochastic differential equations
(SDEs) as follows:
\begin{eqnarray}
dX &=& \sqrt{2\kappa_{xx}}dW_{x}(s)-V_{x}^{sw}ds  \nonumber\\
dY &=& \sqrt{2\kappa_{yy}}dW_{y}(s)-V_{y}^{sw}ds  \nonumber\\
dZ &=& -(\mu V+V_{z}^{sw})ds  \nonumber\\
d\mu &=& \sqrt{2D_{\mu\mu}}dW_{\mu}(s)  \nonumber\\
{}&& -\frac{1-\mu^{2}}{2}\left[\frac{V}{L}+\mu\left(\frac{\partial
V_{x}^{sw}}{\partial x}+\frac{\partial V_{y}^{sw}}{\partial
y}-2\frac{\partial V_{z}^{sw}}{\partial z}\right)\right]ds  \nonumber\\
{}&& +\left(\frac{\partial D_{\mu\mu}}{\partial
\mu}+\frac{2D_{\mu\mu}}{M+\mu}\right)ds  \nonumber\\
dP &=& P\left[\frac{1-\mu^{2}}{2}\left(\frac{\partial
V_{x}^{sw}}{\partial x}+\frac{\partial V_{y}^{sw}}{\partial
y}\right)+\mu^{2}\frac{\partial V_{z}^{sw}}{\partial z}\right]ds,
\label{eq:stochastic-process}
\end{eqnarray}
where $(X,Y,Z)$ is pseudo-position of particles, $V$ is pseudo-speed
of particles, $P$ is pseudo-momentum of particles, and $W_{x}(t)$,
$W_{y}(t)$, and $W_{\mu}(t)$ are so-called Wiener processes. The
five stochastic differential equations represent the value of the
gyrophase-averaged distribution function $f(\textbf{x},\mu,p,t)$ of
SEPs. Using the numerical technique of time-backward Markov
stochastic process, we need to trace SEPs back to the initial time
of the system. The stochastic process simulation starts at the
position $(X,Y,Z)$, pitch-angle $\mu$, momentum $P$, and time $t$
(corresponding to backward time $s=0$), i.e., $X(s=0)=X$,
$Y(s=0)=Y$, $Z(s=0)=Z$, $\mu(s=0)=\mu$, and $P(s=0)=P$. During the
time-backward process, the solution to the gyrophase-averaged
distribution function of particles is sought. All stochastic
simulations exit the system when the trajectories hit the physical
boundaries for the first time. At the initial time, only those
particles reaching the source region can contribute to the
statistics. The five stochastic differential equations are similar
to the first-order ordinary differential equations. In our
simulations, the stochastic differential Equations
(\ref{eq:stochastic-process}) are numerically solved with an Euler
scheme as usual.

The source term $Q$ in Equation (\ref{transport-equation}) is
assumed to be the following form \citep{Reid1964}
\begin{equation}
Q(R\leqslant0.05AU,E_{k},\theta,\phi,t)=\frac{C}{t}\frac{E_{k}^{-\gamma}}{p^{2}}
\exp\left(-\frac{\tau_{c}}{t}-\frac{t}{\tau_{L}}\right)\xi(\theta,\phi),
\label{source}
\end{equation}
where $\gamma$ is spectral index (set to be $3$ in this work) of
source particles, $\tau_{c}$ and $\tau_{L}$ are time constants
controlling particle release profile in SEP sources, and
$\xi(\theta,\phi)$ is a function indicating latitudinal and
longitudinal variation of SEP injection strength in sources. The SEP
source model as shown in Equation (\ref{source}) is usually used to
describe the SEP injections from solar flares in the simulations.
The Equation (\ref{source}) may also be employed to model the
short-lived SEP injections from shock waves driven by coronal mass
ejections (CMEs) near the Sun. This near-Sun injection scenario
should be particularly possible for the high energy particles (e.g.,
$E\gtrsim10$ MeV for protons), which are thought to be accelerated
and released near the Sun, where the shock is quite fast, and the
seed particles are very dense. It should be noted that a
sufficiently strong CME-driven shock can propagate radially outward
to large radial distances, and continue to accelerate particles
during its passage through the solar wind. In this case, the prompt
particle peak can usually be followed by a secondary particle peak,
which is known as energetic storm particle (ESP) event
\citep[e.g.,][]{Verkhoglyadova2012}. In this work, we mainly
concentrate on the high energy SEPs accelerated and released near
the Sun, either from solar flares or from CME-driven shocks in the
corona. We focus on the SEP peak intensities observed in the prompt
component of SEP events. A more complete modeling effort taking into
account the continuous particle injection and the ESP effect
(especially for those relatively low-energy particles) will be the
subject of future work. For a detailed discussion regarding the
effects of the continuous shock acceleration on the SEP flux
profiles, we refer the reader to the work by
\citet{Verkhoglyadova2012}. In addition, the CME intensity increases
usually show an east-west effect, depending on the shock strength
variation with respect to the centroid of the CME and the shock
angle. The long-lived particle injections from such radially
propagating CME-driven shocks may lead to the corotation effect for
spacecraft observations of intensity profiles in the interplanetary
space \citep[e.g.,][]{Lario2014}.

For the outer boundary condition of the numerical model, we use an
absorptive boundary of particles at heliocentric radial distance
$R=50$ AU. In the simulations, we typically use a constant solar
wind speed $V^{sw}=400~km~s^{-1}$ and a Parker-type interplanetary
magnetic field with magnitude $B=5 nT$ at $1 AU$.

In this paper, we pay main attention to the radial dependence of SEP
peak intensities in the inner heliosphere. To this aim, in the
numerical modeling we design two different alignment scenarios of
the fleet of spacecraft in the interplanetary space. As shown in
Figure \ref{SEP-scenarios}, the spacecraft fleet locations labelled
with $A_{i}$ ($i=1, 2, 3, 4, 5$) are aligned along the radial
direction line originating from the SEP source, and the spacecraft
fleet locations labelled with $B_{i}$ ($i=1, 2, 3, 4, 5$) are
aligned along the nominal Parker magnetic field line connecting the
SEP source. The heliocentric radial distances of spacecraft
locations $A_{i}$ (also $B_{i}$) ($i=1, 2, 3, 4, 5$) are in
sequence: $0.25$, $0.4$, $0.6$, $0.8$, and $1.0$ AU. Both the SEP
sources and the fleet of spacecraft are located at $90^{\circ}$
colatitude in this study. Following the energy channels of $27-37$
MeV and $15-40$ MeV protons in SEP events previously analyzed by
\citet{Lario2006} and \citet{Lario2013}, respectively, we
numerically simulate the transport and radial dependence of solar
energetic protons with averaged energies of the energy channels,
i.e., $32$ MeV and $25$ MeV. We use different parallel and
perpendicular diffusion coefficients in the simulations to test the
dependence of the modelling results on these two parameters and on
the ratio of them. In addition, different SEP source coverages
(e.g., $45^{\circ}$ and $70^{\circ}$) of longitude and latitude will
be used in the numerical modelling. We simulate $3\times10^{7}$
particles for each SEP case on a super-computer cluster. The unit of
omnidirectional flux is usually used as
$cm^{-2}-s^{-1}-sr^{-1}-MeV^{-1}$. For conveniently plotting
figures, in this work we use an arbitrary unit.

\section{Numerical Simulation Results}
\begin{figure}
 \epsscale{0.7}
 \plotone{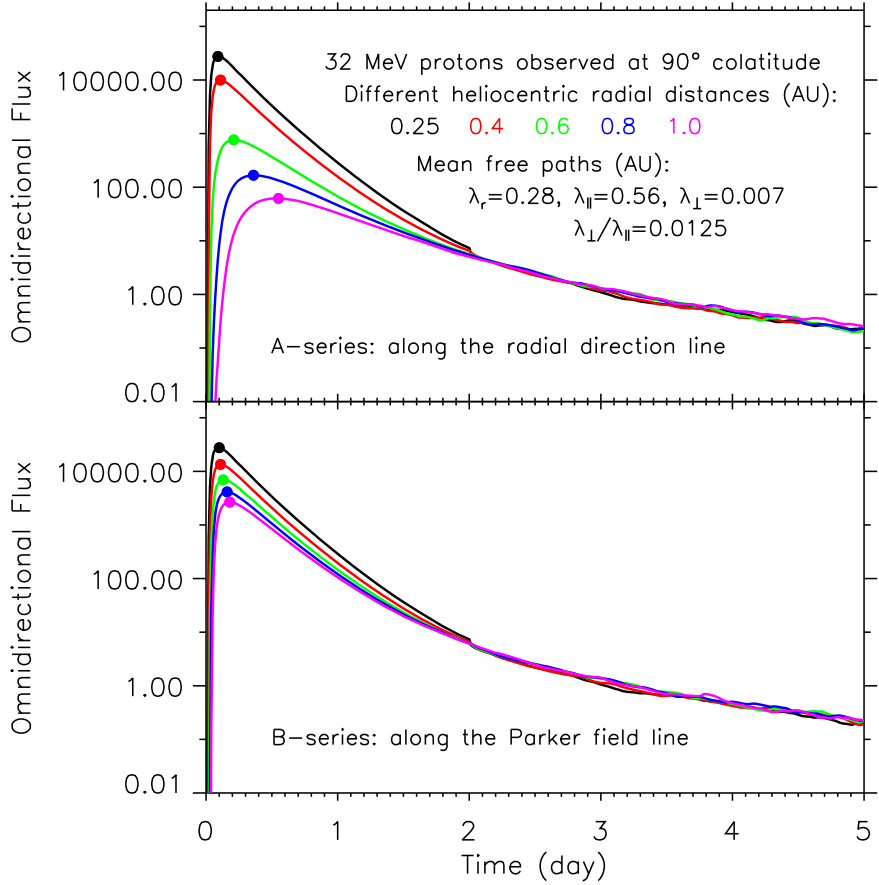}
 \caption{Simulation results of intensity-time profiles of $32$ MeV protons
observed along the radial direction line (upper panel) and along the
Parker magnetic field line (lower panel) originating from the SEP
source. The SEP source is $45^{\circ}$ wide in longitude and
latitude. The diffusion coefficients are set as: the radial mean
free path $\lambda_{r}=0.28$ AU (corresponding to the parallel mean
free path $\lambda_{\parallel}=0.56$ AU at 1 AU), the perpendicular
mean free paths $\lambda_{x}=\lambda_{y}=\lambda_{\perp}=0.007$ AU,
and consequently, the ratio
$\lambda_{\perp}/\lambda_{\parallel}=0.007/0.56=0.0125$ at 1 AU. The
curves with different colors denote the intensity-time profiles
observed at different heliocentric radial distances: $0.25$, $0.4$,
$0.6$, $0.8$, and $1.0$ AU. Note that the SEP ``flood" (previously
``reservoir") phenomenon is reproduced in the simulations of a
series of SEP cases with different radial distances. The filled
circles with different colors on the intensity-time profiles
indicate the peak intensities of the corresponding SEP cases.
\label{32MeV-1}}
\end{figure}

Figure \ref{32MeV-1} shows the simulation results of intensity-time
profiles of $32$ MeV protons observed along the radial direction
line (upper panel) and along the Parker magnetic field line (lower
panel) originating from the SEP source. In the case of Figure
\ref{32MeV-1}, the SEP source coverage is set to be $45^{\circ}$
wide in longitude and latitude. The curves with different colors
denote the intensity-time profiles observed at different
heliocentric radial distances: $0.25$, $0.4$, $0.6$, $0.8$, and
$1.0$ AU. Both the SEP source and the fleet of spacecraft are
located at $90^{\circ}$ colatitude. For both alignment scenarios of
spacecraft (``A-series" and ``B-series" as shown in Figure
\ref{SEP-scenarios}), the diffusion coefficients are set as follows:
the radial mean free path $\lambda_{r}=0.28$ AU (corresponding to
the parallel mean free path $\lambda_{\parallel}=0.56$ AU at 1 AU),
the perpendicular mean free paths
$\lambda_{x}=\lambda_{y}=\lambda_{\perp}=0.007$ AU, and
consequently, the ratio
$\lambda_{\perp}/\lambda_{\parallel}=0.007/0.56=0.0125$ at 1 AU. As
we can see in the upper and lower panels of Figure \ref{32MeV-1},
for both alignment scenarios of spacecraft, the SEP intensity
detected at a smaller radial distance is higher than the SEP
intensity observed at a larger radial distance. Interestingly,
during the early phases of the SEP events, the magnitude difference
between the SEP intensities observed at different radial distances
in the upper panel (``A-series") is much more considerable than that
in the lower panel (``B-series"). The essential reason is that in
the simulations and also in typical SEP events, the perpendicular
diffusion coefficient is smaller than the parallel diffusion
coefficient. As a result, with the same radial distance, the SEP
intensity observed along the radial direction line (``A-series") is
smaller than that observed along the Parker magnetic field line
(``B-series"), especially at larger radial distances, where the
magnetic footpoint of the observer in the ``A-series" cases is
outside the limited SEP source. In the upper and lower panels of
Figure \ref{32MeV-1}, the filled circles with different colors on
the intensity-time profiles indicate the peak intensities of the
corresponding SEP cases. During the late phases of SEP events, the
particle intensities usually evolve in time with similar decay
rates. This interesting phenomenon was traditionally named SEP
``reservoir" by \citet{Roelof1992} and twenty-five years later was
recently tentatively renamed SEP ``flood" by \citet{He2017} based on
multi-dimensional numerical simulations and spacecraft observations.
As one can see, in both panels of Figure \ref{32MeV-1}, the SEP
``flood" (previously ``reservoir") phenomenon is successfully
reproduced in our simulations of a series of SEP cases with
different radial distances.

\begin{figure}
 \epsscale{0.7}
 \plotone{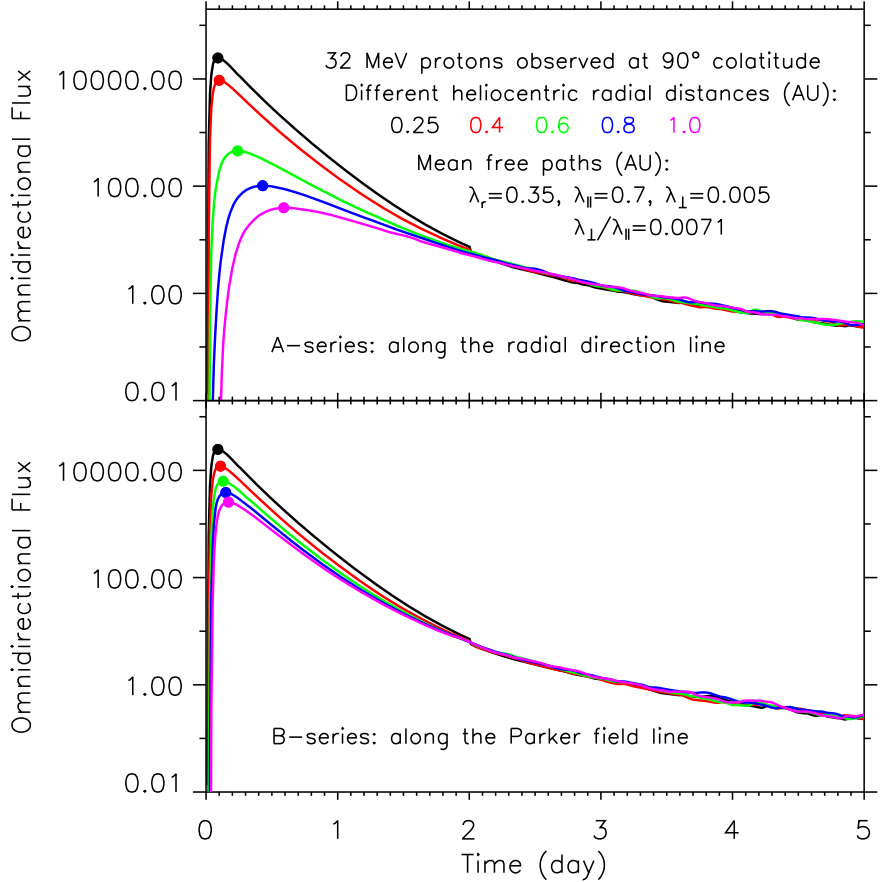}
 \caption{Same as Figure \ref{32MeV-1} except the diffusion coefficients are
set as: the radial mean free path $\lambda_{r}=0.35$ AU
(corresponding to the parallel mean free path
$\lambda_{\parallel}=0.7$ AU at 1 AU), the perpendicular mean free
paths $\lambda_{x}=\lambda_{y}=\lambda_{\perp}=0.005$ AU, and
consequently, the ratio
$\lambda_{\perp}/\lambda_{\parallel}=0.005/0.7=0.0071$ at 1 AU. The
SEP ``flood" (previously ``reservoir") phenomenon is reproduced in
the simulations. \label{32MeV-2}}
\end{figure}

Figure \ref{32MeV-2} shows the simulation results of intensity-time
profiles of $32$ MeV protons with the diffusion coefficients as
follows: the radial mean free path $\lambda_{r}=0.35$ AU
(corresponding to the parallel mean free path
$\lambda_{\parallel}=0.7$ AU at 1 AU), the perpendicular mean free
paths $\lambda_{x}=\lambda_{y}=\lambda_{\perp}=0.005$ AU, and
consequently, the ratio
$\lambda_{\perp}/\lambda_{\parallel}=0.005/0.7=0.0071$ at 1 AU.
Other parameters and conditions are the same as Figure
\ref{32MeV-1}. In Figure \ref{32MeV-2}, we can see that the SEP
intensity observed at a smaller radial distance is higher than that
observed at a larger radial distance. In other words, the SEP
intensity decreases with increasing radial distance. The filled
circles with different colors on the intensity-time profiles
indicate the peak intensities of the SEP events. In addition, the
SEP ``flood" (previously ``reservoir") phenomenon is reproduced in
the simulations.

\begin{figure}
 \epsscale{0.7}
 \plotone{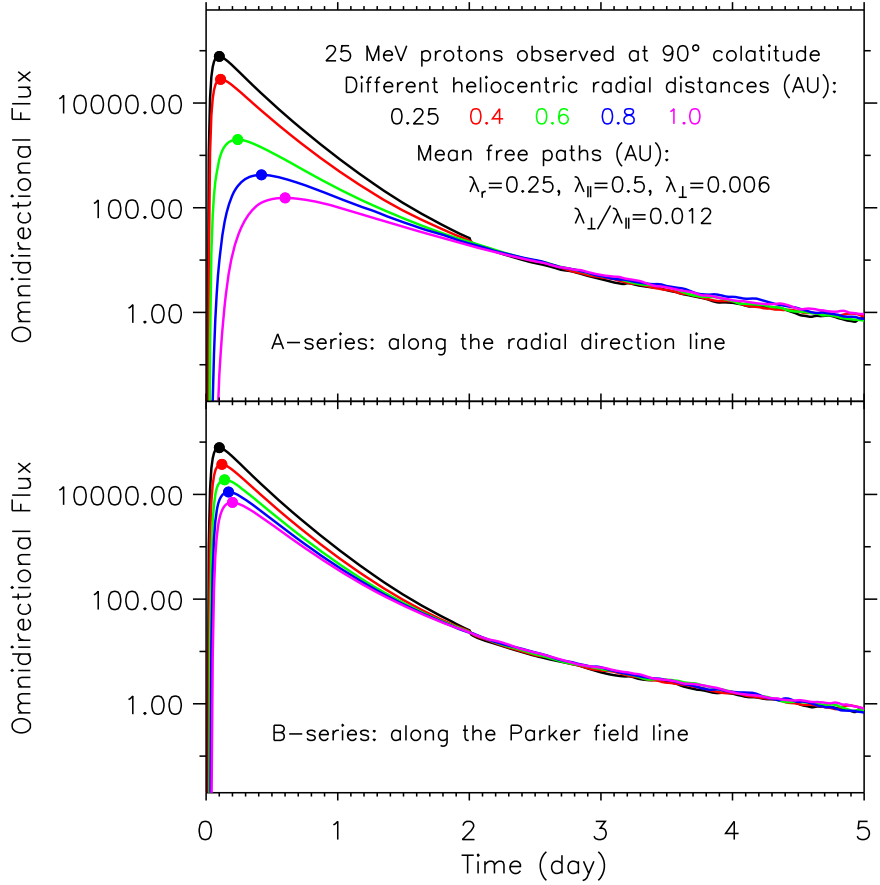}
 \caption{Same as Figure \ref{32MeV-1} except for $25$ MeV protons and the
diffusion coefficients are set as: the radial mean free path
$\lambda_{r}=0.25$ AU (corresponding to the parallel mean free path
$\lambda_{\parallel}=0.5$ AU at 1 AU), the perpendicular mean free
paths $\lambda_{x}=\lambda_{y}=\lambda_{\perp}=0.006$ AU, and
consequently, the ratio
$\lambda_{\perp}/\lambda_{\parallel}=0.006/0.5=0.012$ at 1 AU. The
SEP ``flood" (previously ``reservoir") phenomenon is reproduced in
the simulations. \label{25MeV-1}}
\end{figure}

Figure \ref{25MeV-1} displays the simulation results of
intensity-time profiles of $25$ MeV protons with the diffusion
coefficients as: the radial mean free path $\lambda_{r}=0.25$ AU
(corresponding to the parallel mean free path
$\lambda_{\parallel}=0.5$ AU at 1 AU), the perpendicular mean free
paths $\lambda_{x}=\lambda_{y}=\lambda_{\perp}=0.006$ AU, and
consequently, the ratio
$\lambda_{\perp}/\lambda_{\parallel}=0.006/0.5=0.012$ at 1 AU. Other
parameters and conditions are the same as Figure \ref{32MeV-1}.
Similar to the results shown in Figure \ref{32MeV-1} and Figure
\ref{32MeV-2}, we can see that for both alignment scenarios of
spacecraft (``A-series" in upper panel and ``B-series" in lower
panel), the SEP intensity decreases with increasing radial distance.
The filled circles on the intensity-time profiles denote the peak
intensities of the SEP events. During the late phases of SEP events,
the SEP ``flood" (previously ``reservoir") phenomenon is clearly
reproduced.

\begin{figure}
 \epsscale{0.7}
 \plotone{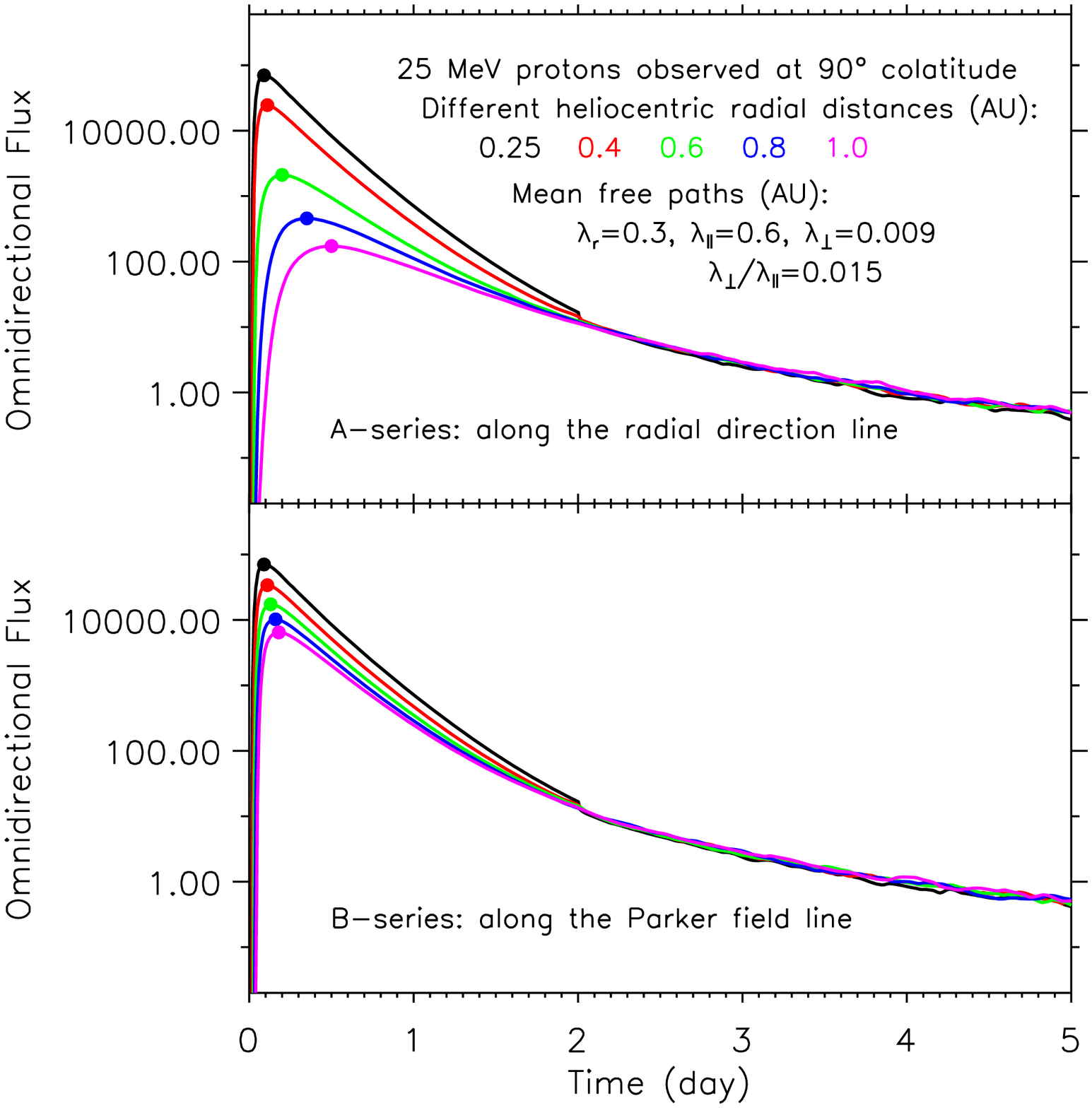}
 \caption{Same as Figure \ref{25MeV-1} except the diffusion coefficients are
set as: the radial mean free path $\lambda_{r}=0.3$ AU
(corresponding to the parallel mean free path
$\lambda_{\parallel}=0.6$ AU at 1 AU), the perpendicular mean free
paths $\lambda_{x}=\lambda_{y}=\lambda_{\perp}=0.009$ AU, and
consequently, the ratio
$\lambda_{\perp}/\lambda_{\parallel}=0.009/0.6=0.015$ at 1 AU.
During the late phases, the SEP ``flood" (previously ``reservoir")
phenomenon is reproduced. \label{25MeV-2}}
\end{figure}

Figure \ref{25MeV-2} shows the numerical simulation results of
intensity-time profiles of $25$ MeV protons with the diffusion
coefficients as: the radial mean free path $\lambda_{r}=0.3$ AU
(corresponding to the parallel mean free path
$\lambda_{\parallel}=0.6$ AU at 1 AU), the perpendicular mean free
paths $\lambda_{x}=\lambda_{y}=\lambda_{\perp}=0.009$ AU, and
consequently, the ratio
$\lambda_{\perp}/\lambda_{\parallel}=0.009/0.6=0.015$ at 1 AU. Other
parameters and conditions are the same as Figure \ref{25MeV-1}. As
one can see in Figure \ref{25MeV-2}, the SEP intensity decreases
with increasing radial distance. We note that the filled circles on
the intensity-time profiles denote the peak intensities of the
corresponding SEP cases. During the late phases, the SEP ``flood"
(previously ``reservoir") phenomenon is reproduced.

\begin{figure}
 \epsscale{1.0}
 \plotone{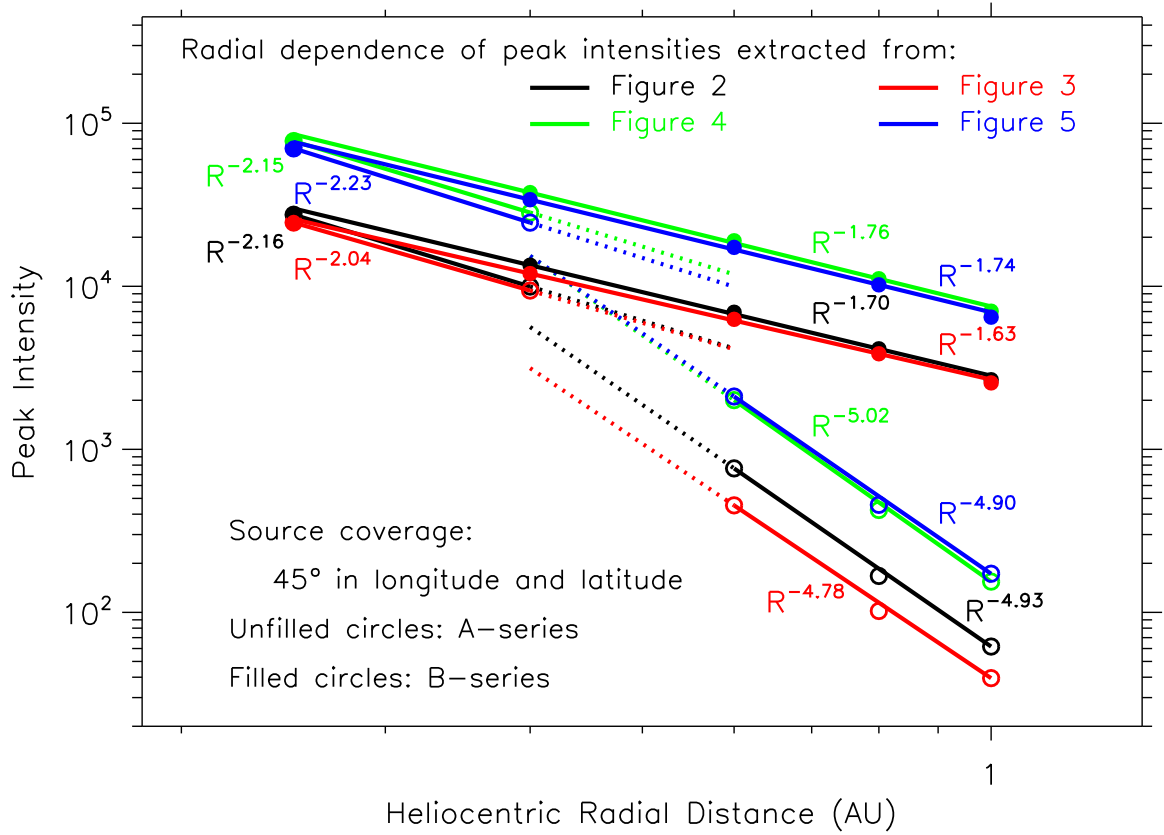}
 \caption{Radial dependence of the SEP peak intensities extracted from Figures
\ref{32MeV-1} (black), \ref{32MeV-2} (red), \ref{25MeV-1} (green),
and \ref{25MeV-2} (blue). The unfilled circles denote the peak
intensities of the SEP events observed along the radial direction
line, and the filled circles denote the peak intensities of the SEP
events observed along the Parker magnetic field line. Note that the
SEP source is $45^{\circ}$ wide in longitude and latitude. The
functional form $R^{-\alpha}$ with different values of $\alpha$
denotes the modelling results of the corresponding SEP cases. A
two-phase phenomenon is found in the radial dependence of the SEP
peak intensities. The ``position" of the breakpoint between the two
phases depends on the magnetic connection status of the observers.
\label{intensity-radial}}
\end{figure}

We extract the information of the peak intensities and the relevant
radial distances of the SEP cases in Figures \ref{32MeV-1},
\ref{32MeV-2}, \ref{25MeV-1}, and \ref{25MeV-2}. We present this
important information in Figure \ref{intensity-radial}. The black,
red, green, and blue circles and curves in Figure
\ref{intensity-radial} denote the simulation results of SEP peak
intensities extracted from Figures \ref{32MeV-1}, \ref{32MeV-2},
\ref{25MeV-1}, and \ref{25MeV-2}, respectively. In previous
observational studies, a power-law function
$I_{max}(R)=kR^{-\alpha}$, where $k$ is a constant, is usually used
to model the radial dependence of the SEP peak intensities.
Following the previous works, we model the radial dependence of the
SEP peak intensities shown in Figures \ref{32MeV-1}, \ref{32MeV-2},
\ref{25MeV-1}, and \ref{25MeV-2} with the functional form
$I_{max}(R)=kR^{-\alpha}$. In Figure \ref{intensity-radial}, the
unfilled circles denote the peak intensities of the ``A-series" SEP
cases, i.e., those events observed along the radial direction line.
The filled circles denote the peak intensities of the ``B-series"
SEP cases, i.e., those events observed along the Parker magnetic
field line. The functional form $R^{-\alpha}$ with different values
of $\alpha$ near the corresponding SEP cases denotes the modelling
results of the radial dependence of the SEP peak intensities. As we
can see, the peak particle intensities in the SEP events
exponentially decrease with the increasing radial distances. For the
SEP events observed along the Parker magnetic field line
(``B-series"), the values of $\alpha$ are roughly in the range
$[1.6, 1.8]$, with a median value of $1.7$. For the SEP events
observed along the radial direction line (``A-series"), the values
of $\alpha$ are generally larger than those in the ``B-series" SEP
events. Interestingly, a two-phase phenomenon is found in the radial
dependence of the SEP peak intensities. For the two spacecraft at
radial distances $R=0.25$ AU and $R=0.4$ AU, respectively, the
values of $\alpha$ are roughly in the range $[2.0, 2.2]$. For the
three spacecraft at radial distances $R=0.6$ AU, $R=0.8$ AU, and
$R=1.0$ AU, respectively, the values of $\alpha$ are roughly in the
range $[4.8, 5.0]$. We note that the magnetic footpoints of the
former two spacecraft are located inside the SEP source, while the
magnetic footpoints of the latter three spacecraft are located
outside the SEP source. This is why the value range $[2.0, 2.2]$ of
$\alpha$ determined from the former two spacecraft is relatively
similar to the value range $[1.6, 1.8]$ of $\alpha$ in the
``B-series" SEP cases, but the value range $[4.8, 5.0]$ of $\alpha$
determined from the latter three spacecraft is considerably
different from the value range $[1.6, 1.8]$ of $\alpha$ in the
``B-series" SEP cases, where all of the five spacecraft's magnetic
footpoints are located at the center of the SEP source. The
``position" of the breakpoint (transition point/critical point)
between the two phases of the radial dependence of the SEP peak
intensities depends on the magnetic connection status of the
observers. In particular, the situation that whether the magnetic
footpoint of the observer is located inside (even very near or on
the boundary of source) or outside the SEP source is a crucial
factor determining the value of the index $\alpha$. This finding
suggests that a very careful examination of magnetic connection
between SEP source and each spacecraft should be taken in the
observational investigations to obtain an accurate estimate of the
radial dependence of the SEP peak intensities. In addition, we find
that the index $\alpha$ does not strongly depend on the energies of
particles and the ratios of perpendicular to parallel diffusion
coefficients, provided that the diffusion coefficients are
relatively reasonable.

\begin{figure}
 \epsscale{1.0}
 \plotone{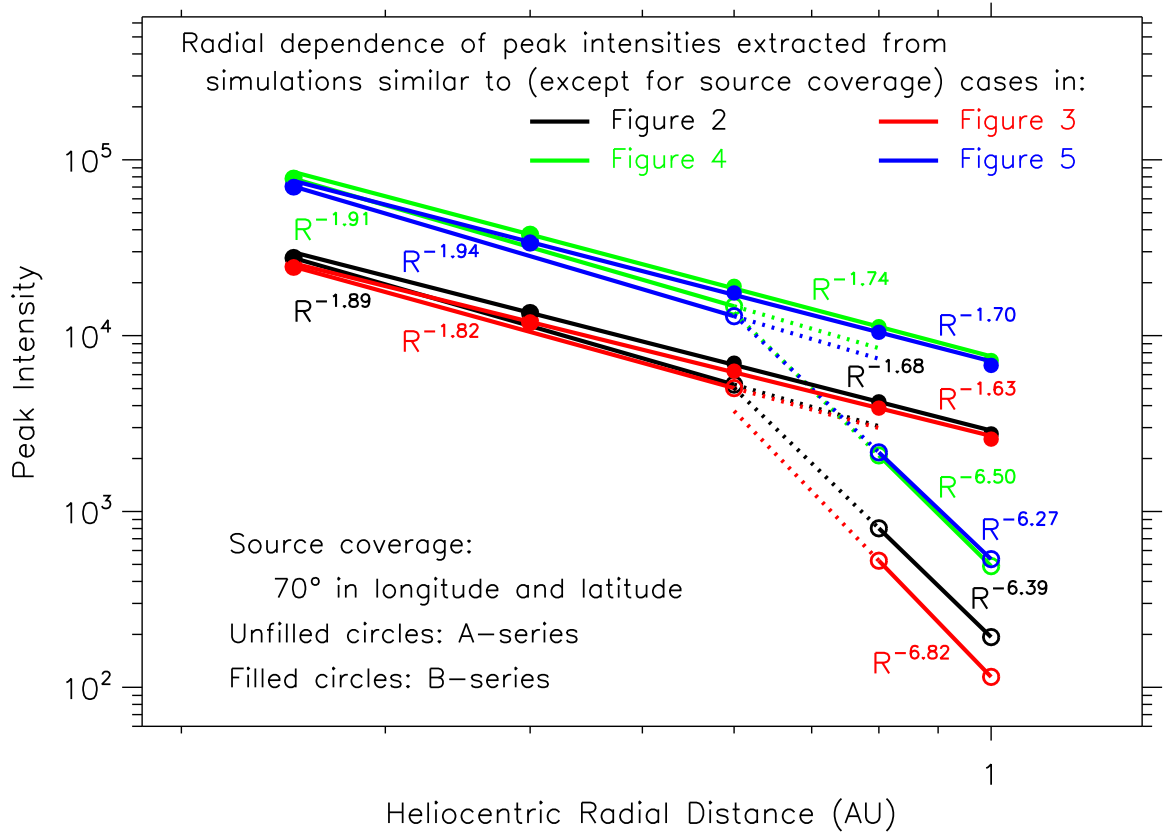}
 \caption{Same as Figure \ref{intensity-radial} except for the SEP source
being $70^{\circ}$ wide in longitude and latitude. A two-phase
phenomenon also exists in the radial dependence of the SEP peak
intensities. The ``position" of the breakpoint between the two
phases depends on the magnetic connection status of the observers.
\label{intensity-radial-big}}
\end{figure}

We also simulate the SEP cases with source width of $70^{\circ}$ in
longitude and latitude. Figure \ref{intensity-radial-big} displays
the simulation results and the modelling analyses of the radial
dependence of the SEP peak intensities. The black, red, green, and
blue circles and curves in Figure \ref{intensity-radial-big} denote
the simulation results of the SEP cases similar to (except for the
source coverage) the cases presented in Figures \ref{32MeV-1},
\ref{32MeV-2}, \ref{25MeV-1}, and \ref{25MeV-2}, respectively. We
note that except for the source width, all of the other physical
parameters and conditions of the simulations presented in Figure
\ref{intensity-radial-big} are the same as those in Figure
\ref{intensity-radial}. The unfilled circles denote the peak
intensities of the ``A-series" SEP cases, and the filled circles
denote the peak intensities of the ``B-series" SEP cases. The
functional form $R^{-\alpha}$ with different values of $\alpha$ near
the corresponding SEP cases denotes the modelling results of the
radial dependence of the SEP peak intensities. For the SEP events
observed along the Parker magnetic field line (``B-series"), the
values of $\alpha$ are roughly in the range $[1.6, 1.7]$, which is
similar to the value range $[1.6, 1.8]$ indicated in Figure
\ref{intensity-radial}. For the SEP events observed along the radial
direction line (``A-series"), the values of $\alpha$ are generally
larger than those in the ``B-series" SEP events. Similar to Figure
\ref{intensity-radial}, a two-phase phenomenon exists in the radial
dependence of the SEP peak intensities. For the three spacecraft at
radial distances $R=0.25$ AU, $R=0.4$ AU, and $R=0.6$ AU,
respectively, the values of $\alpha$ are roughly in the range $[1.8,
1.9]$. For the two spacecraft at radial distances $R=0.8$ AU and
$R=1.0$ AU, respectively, the values of $\alpha$ are roughly in the
range $[6.3, 6.8]$. We note that the magnetic footpoints of the
former three spacecraft are located inside the SEP source, while the
magnetic footpoints of the latter two spacecraft are located outside
the SEP source. This is why the value range $[1.8, 1.9]$ of $\alpha$
determined from the former three spacecraft is similar to the value
range $[1.6, 1.7]$ of $\alpha$ in the ``B-series" SEP cases, but the
value range $[6.3, 6.8]$ of $\alpha$ determined from the latter two
spacecraft is considerably distinct from the value range $[1.6,
1.7]$ of $\alpha$ in the ``B-series" SEP cases. Because for the
``B-series" SEP cases, the magnetic footpoints of all of the five
spacecraft are located at the center of the SEP source. By combining
Figure \ref{intensity-radial} and Figure \ref{intensity-radial-big},
we can conclude that the ``position" of the breakpoint (transition
point/critical point) between the two phases of the radial
dependence of the SEP peak intensities depends on the magnetic
connection status of the observers in the heliosphere. Particularly,
the situation that whether the magnetic footpoint of each observer
is located inside (even very near or on the boundary of source) or
outside the SEP source is a very important factor determining the
value of the index $\alpha$. In this sense, the index $\alpha$
primarily depends on the physical properties (e.g., location,
coverage) of the SEP sources and consequently on the
longitudinal/latitudinal separations between the sources and the
magnetic field line footpoints of the observers. Therefore, we
suggest that a very careful examination of the magnetic connection
between SEP source and each spacecraft should be taken in the
observational studies to obtain an accurate estimate of the radial
dependence of the SEP peak intensities. Additionally, we find that
the index $\alpha$ does not strongly depend on the energies of
particles and the ratios of perpendicular to parallel diffusion
coefficients. In addition, we suggest that the function
$R^{-1.7\pm0.1}$, obtained in the ``B-series" SEP cases, where all
of the observers' magnetic footpoints are located at the center of
the SEP source, should be the lower limit for empirically modelling
the radial dependence of the SEP peak intensities. For the SEP
events similar to the ``B-series" cases as shown in Figure
\ref{SEP-scenarios}, the value of index $\alpha$ does not strongly
depend on the width of the SEP sources. These findings can be used
to explain the majority of the previous empirical investigation
results based on multispacecraft observations, and especially to
reconcile the different/conflicting empirical values of index
$\alpha$ in the literature.

\section{Summary and Conclusions}
\begin{figure}
 \epsscale{1.0}
 \plotone{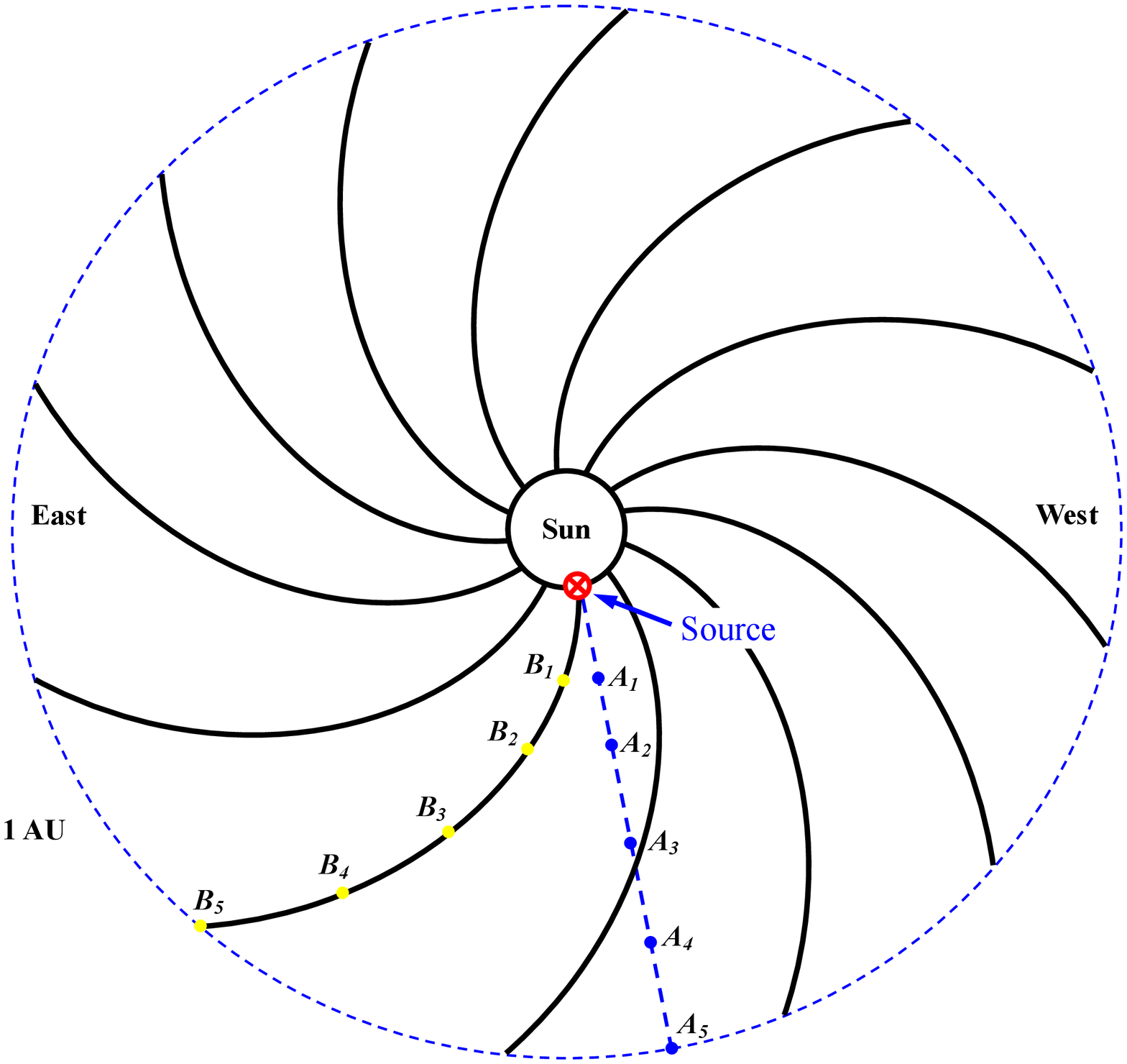}
 \caption{Diagram to show the alignment scenarios of the fleet of spacecraft
in the heliosphere. The spacecraft fleet locations labelled with
$A_{i}$ ($i=1, 2, 3, 4, 5$) are aligned along the radial direction
line, and the spacecraft fleet locations labelled with $B_{i}$
($i=1, 2, 3, 4, 5$) are aligned along the Parker magnetic field line
connecting the SEP source. The heliocentric radial distances of
spacecraft locations $A_{i}$ (also $B_{i}$) ($i=1, 2, 3, 4, 5$) are
in sequence: $0.25$, $0.4$, $0.6$, $0.8$, and $1.0$ AU. Both the SEP
sources and the fleet of spacecraft are located at $90^{\circ}$
colatitude. \label{SEP-scenarios}}
\end{figure}

SEPs have radiation effects and potential damages on space missions,
and can risk the health of the astronauts working in space. To
accurately estimate the potential radiation impacts of SEPs, it is
very important to investigate the radial dependence of SEP
intensities and fluences, particularly in the forthcoming era of
Solar Probe Plus and Solar Orbiter. In this paper, we numerically
solve the five-dimensional Fokker-Planck equation to investigate the
radial dependence of SEP peak intensities in the inner heliosphere.
The value of index $\alpha$ in the functional form
$I_{max}(R)=kR^{-\alpha}$ is quantitatively determined in the
simulations of various SEP scenarios. Two different styles of
spacecraft alignment in the heliosphere are taken into account: (1)
along the radial direction line (``A-series" cases in Figure
\ref{SEP-scenarios}); (2) along the nominal Parker magnetic field
line (``B-series" cases in Figure \ref{SEP-scenarios}). The main
conclusions of our results in this paper are as follows:

1. The value of index $\alpha$ varies in a wide range, mainly
depending on the physical properties (e.g., location, width) of SEP
sources and consequently on the longitudinal/latitudinal separations
between the sources and the magnetic field line footpoints of the
observers. However, for the SEP events similar to the ``B-series"
cases, the value of $\alpha$ does not strongly depend on the width
of the SEP sources.

2. The situation that whether the magnetic footpoint of the observer
is located inside (even very near or on the boundary of source) or
outside the SEP source is a crucial factor determining the value of
index $\alpha$. A two-phase phenomenon is found in the radial
dependence of the peak intensities in the ``A-series" SEP cases. The
``position" of the breakpoint (transition point/critical point)
between the two phases of the radial dependence of the peak
intensities depends on the magnetic connection status of the
observers. We suggest that a very careful examination of the
magnetic connection between SEP source and each spacecraft should be
taken in the observational investigations to obtain an accurate
estimate of the radial dependence of the SEP peak intensities.

3. We suggest that the function $R^{-1.7\pm0.1}$, obtained in the
``B-series" SEP cases, where all of the observers' magnetic
footpoints are located at the center of the SEP source, should be
the lower limit for empirically modelling the radial dependence of
the SEP peak intensities. In addition, the value of the index
$\alpha$ does not strongly depend on the energies of particles and
the ratios of perpendicular to parallel diffusion coefficients,
provided that the diffusion coefficients are relatively reasonable.

4. The results provided in this paper can be used to explain the
majority of the previous empirical results based on multispacecraft
observations, and especially to reconcile the different/conflicting
empirical values of index $\alpha$ in the literature. In addition,
our findings can also be employed to predict and explain the
observational results obtained from the two forthcoming spacecraft,
i.e., Solar Probe Plus and Solar Orbiter.

In this work, we primarily pay attention to the simulations of the
three-dimensional propagation of energetic protons in the inner
heliosphere. In the future, we will investigate the radial
dependence of intensities of electrons and heavy ions. The radial
dependence of SEP fluences and radiation dosages is also a very
interesting topic in the research field. We will also investigate
the radial dependence of SEP intensities, SEP fluences, and
radiation dosages in the outer heliosphere. In addition, in the
future we will further investigate the radial and temporal evolution
of the SEP events related to CME-driven shocks.

%%%%%%%%%%%%%%%%%%%%%%%%%%%%%%%%%%%%%%%%%%%%%%%%%%%%%%%%%%%%%%%%%

\acknowledgments

This work was supported in part by the National Natural Science
Foundation of China under grants 41474154, 41204130, 41621063, and
41131066, the National Important Basic Research Project under grant
2011CB811405, the Chinese Academy of Sciences under grant
KZZD-EW-01-2. H.-Q. He gratefully acknowledges the partial support
of the Youth Innovation Promotion Association of the Chinese Academy
of Sciences (No. 2017091).

%%%%%%%%%%%%%%%%%%%%%%%%%%%%%%%%%%%%%%%%%%%%%%%%%%%%%%%%%%%%%%%%%

\clearpage

%%%%%%%%%%%%%%%%%%%%%%%%%%%%%%%%%%%%%%%%%%%%%%%%%%%%%%%%%%%%%%%%%

\clearpage

\clearpage

\clearpage

\clearpage

\clearpage

\clearpage

\clearpage

%%%%%%%%%%%%%%%%%%%%%%%%%%%%%%%%%%%%%%%%%%%%%%%%%%%%%%%%%%%%%%%%%

\end{document}